\documentclass[a4paper]{jpconf}

\usepackage{graphicx}
\usepackage{wrapfig}
\usepackage{textgreek}

\begin{document}

\title{Recent advancements of the NEWS-G experiment}

\author{Ioannis Katsioulas\\on behalf of the NEWS-G collaboration}

\address{School of Physics and Astronomy, University of Birmingham, B15 2TT, United Kingdom}

\ead{i.katsioulas@bham.ac.uk}

\begin{abstract}
NEWS-G (New Experiments With Spheres-Gas) is an experiment aiming to shine a light on the dark matter conundrum with a novel gaseous detector, the spherical proportional counter. It uses light gases, such as hydrogen, helium, and neon, as targets to expand dark matter searches to the sub-GeV/c$^{2}$ mass region. NEWS-G produced its first results with a 60 cm in diameter detector installed at LSM (France), excluding at 90\% C.L. cross-sections above $4.4\cdot{10}^{37}$ cm$^{2}$ for dark matter candidates of 0.5 GeV/c$^{2}$ mass. Currently, a 140 cm in diameter detector is being built at LSM and a commissioning run is underway, prior to its installation at SNOLAB (Canada) at the end of the year. Presented here are developments incorporated in this new detector: a) sensor technologies using resistive materials and multi-anode read-out that allow high gain and high pressure operation; b) gas purification techniques to remove contaminants (H$_{2}$O, O$_{2}$); c) reduction of ${}^{210}$Pb induced background through copper electroforming methods; d) utilisation of UV-lasers for detector calibration, detector response monitoring and estimation of gas related fundamental properties. This next phase of NEWS-G will allow searches for low mass dark matter with unprecedented sensitivity.
\end{abstract}

\section{Introduction}

The NEWS-G collaboration is searching for light dark matter (DM) \cite{dm} where many new theoretical approaches such as the asymmetric dark model and dark sector predict DM candidates. These searches are performed using a gaseous particle detector, the Spherical Proportional Counter (SPC) \cite{spc_1} filled with light mass gases \cite{news_idea}, such as neon, methane, and helium.

\begin{wrapfigure}{r}{0.4\textwidth}
\centering
\includegraphics[width=0.38\textwidth]{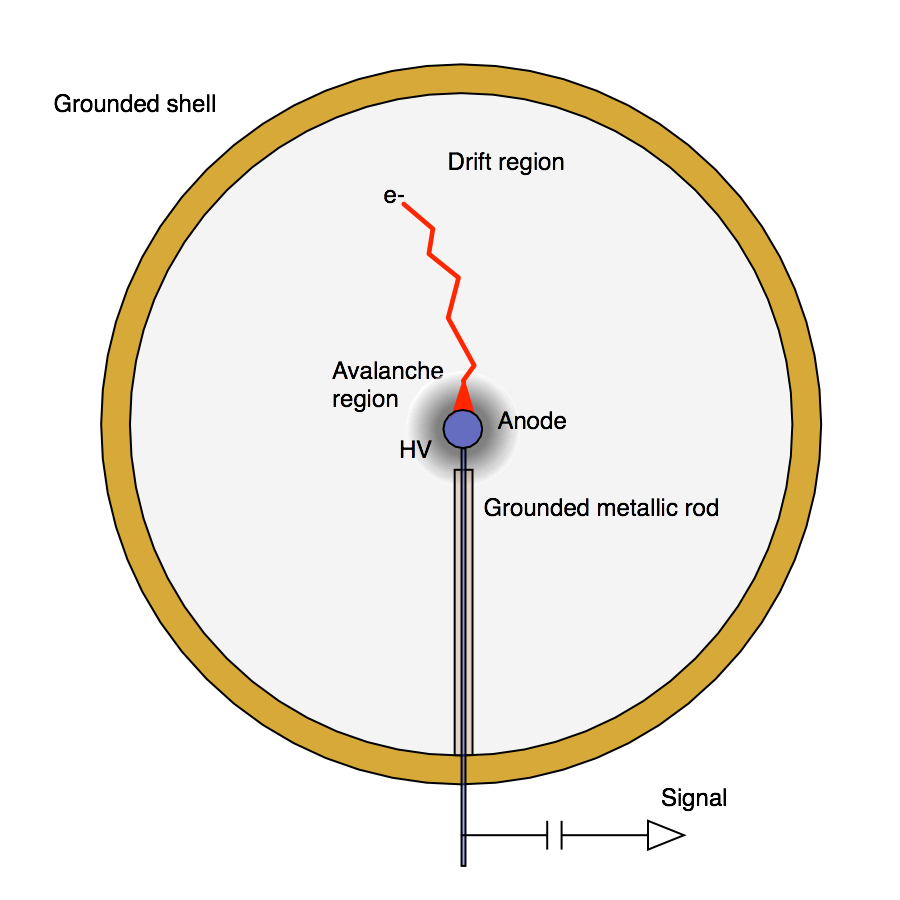}
\caption{\label{spc_draw} SPC design and principle of operation.}
\end{wrapfigure}

The SPC consists of a grounded metallic spherical shell shown in Fig. \ref{spc_draw}. A small sensor is placed at the center of the sphere supported by a grounded metallic rod, and is held at positive high voltage. The resulting electric field is mostly radial, except near the sensor rod which disturbs the field, and falls as 1/r$^{2}$.  The low capacitance of the sensor, which allows for low electronic noise, in combination with the large amplification of the signal, provides single electron detection and therefore makes the SPC a powerful detector for low energy nuclear recoils.
NEWS-G operated its prototype, SEDINE a 60-cm diameter SPC made from pure copper at the Laboratoire Souterrain de Modane (LSM), in France, primarily to prove the concept of using large SPCs to search for low-mass dark matter. To further mitigate background from external radiation, SEDINE was put inside a multi-layered cubic shielding composed of, from the inside to the outside, 8 cm of copper, 15 cm of lead and 30 cm of polyethylene. At the center of SEDINE, a grounded copper rod is holding a 6.3-mm silicon sensor held at high-voltage. Between April and May 2015, while it was filled to a pressure of 3.1  bar with a mixture of 90.3\% neon and 0.7\% methane, SEDINE ran in dark matter search mode uninterruptedly for 42.7 days. In 2017 NEWS-G set new constraints on the spin-independent WIMP-nucleon scattering cross-section below 0.6 GeV/c$^{2}$ and excluded at 90\% C.L. a cross-section of $4.4\cdot {10}^{−37}$ cm$^{2}$ for a 0.5 GeV/c$^{2}$ light DM candidate mass. The details for the experimental setup, result, pulse treatment, data analysis, were shown in \cite{first}. 

The NEWS-G collaboration is planning to install of a 140-cm diameter SPC made from ultra-pure copper (commercial C10100) (figure 7). This detector is the largest SPC to date and has been approved for installation at SNOLAB in Canada. The design of the shielding is much more advanced and compact than SEDINE's, comprised from the outside to the inside, 40 cm of borated polyethylene, and 22 cm of low activity lead (including the inner most 3 cm made from archaeological lead). The lead shield is placed into a stainless steel envelope that will be flushed with pure nitrogen to mitigate the presence of radon. All of the shielding and detector will be sitting on a seismic platform as a precaution for seismic events. 
A commissioning of the experiment took place during summer 2019 at LSM, while the polyethylene shielding is being fabricated in Canada. The construction of a neutron shielding based on a concentric cylindrical water tank at LSM will allow for a first short dark matter search run, before the detector was shipped to SNOLAB. It is expected that the dark matter search at SNOLAB will begin in Winter 2020.

\section{ACHINOS - The multi-anode SPC sensor}

\begin{figure}[h]
\centering
\begin{minipage}{15pc}
\centering
\includegraphics[width=12pc]{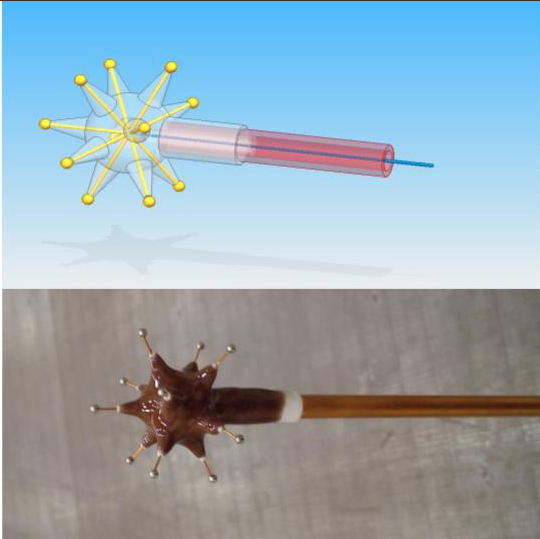}
\caption{\label{ach_des}The design of an 11-anode ACHINOS and the implementation covering the 3D printed central electrode with a resistive paste.}
\end{minipage}\hspace{2pc}%
\begin{minipage}{16pc}
\includegraphics[width=16pc]{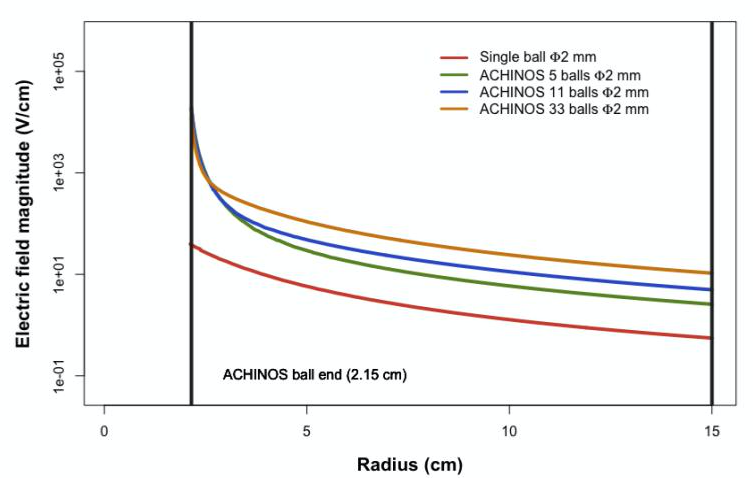}
\caption{\label{ach_res} 
Magnitude of electric field of ACHINOS sensors with 5, 11, and 33 anodes compared to the electric field of a single ball sensor with its anode in the center of the detector, and in the same potential \cite{ach_1}.}
\end{minipage} 
\end{figure}

The ACHINOS (Greek for sea-urchin) multi-anode sensor consists of a set of anode balls uniformly distributed around a central sphere at an equal distance from the center of the detector, supported by insulated wires through which HV can be applied (HV1) on them. The central sphere is used as a bias electrode by applying (HV2) on its surface which helps to optimise the electric field configuration. The example of an 11-anode ACHINOS is displayed in Fig. \ref{ach_des}. The motivation for the development of such an instrument is to provide an increased electric field in the large radii of large spherical proportional counters which in the case of single anode sensors can be below 0.1 V/cm. By increasing the electric field magnitude electrons and ions are collected faster making operation of the detector less sensitive to attachment induced by O$_{2}$, H$_{2}$ presence. This task its achieved with ACHINOS without high gain operation burdened by using low diameter anodes (below 2 mm) and in the same time increase the strength of the electric field in the detector volume by increasing the number of anodes and their distance from the central secondary electrode.  The effect of multiple anodes being used in an ACHINOS sensor versus a single anode sensor is displayed in figure Fig. \ref{ach_res}. The electric field close to the surface of the shell of the detector is higher in the case of an ACHINOS sensor (approximately 9 times higher for an 11-ball ACHINOS) than in the case of the single ball sensor, an effect depicted in the measured reduction of the maximum risetime of pulses \cite{ach_1}.

\newpage

\section{Gas purification}

\begin{wrapfigure}{r}{0.4\textwidth}
\centering
\includegraphics[width=0.39\textwidth]{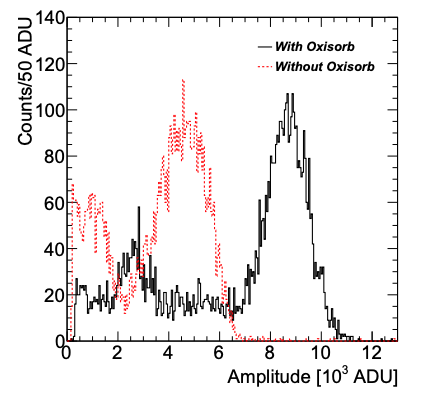}
\caption{\label{filter} Signal comparison of 5.9 keV X-rays in a spherical proportional counter filled with 600 mbar of He:CH$_{4}$ (90\%:10\%) gas with and without filtering \cite{filter}.}
\end{wrapfigure}

Gas contaminants with electronegative molecules lead in signal reduction, degradation of the energy resolution and background discrimination capabilities. Their effect particularly pronounced in regions of low electric field. Gas filtering using Messer Oxisorb or Saes MicroTorr Purifier was introduced to ensure that oxygen and water induced effects were minimised. Fig. \ref{filter} shows the pulse amplitude for 5.9 keV photons measured with a spherical proportional counter filled with filtered and unfiltered gas. Filtering improved recorded resolution (\textsigma/E), from 21.3 $\pm$ 0.7\% to 9.4 $\pm$ 0.3\%. During the process, it was found that non-negligible amounts of $^{222}$Rn were introduced. This has previously been reported by several experiments \cite{rn} and work is ongoing to incorporate a carbon filter, inserted after the oxygen filter in the gas system to remove any emanated $^{222}$Rn. 

\section{Pulsed laser for detector monitoring\\ and calibration}
 \begin{figure}[h]
\centering
\includegraphics[width=0.9\textwidth]{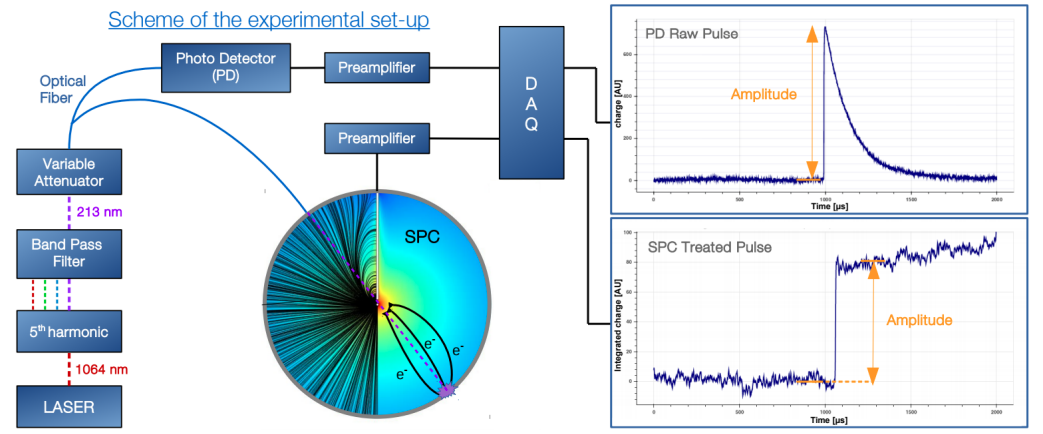}
\caption{\label{laser}Schematic of laser-based calibration showing the principle of operation and example pulses \cite{laser}.  }
\end{figure}

The NEWS-G collaboration has recently reported on a novel precision laser-based calibration that allowed for the measurement of the single-electron-response (SER) in SPCs. A monochromatic UV laser beam with a variable intensity was used to extract single photo-electrons from the cathode of the SPC. The SPC data acquisition is triggered using the laser signal in a photo detector. This allows for the precise measurement of electron transport parameters such as drift time, diffusion coefficients, and electron avalanche gain. A schematic of the experimental setup is shown in \ref{laser}. These studies are complement with an internal $^{37}$Ar source calibration for measurements of the gas W-value and Fano factor. The calibration system can be used during the dark matter search to monitor the detector response. Additionally, the trigger efficiency can be measured using events triggered by the laser photo detector. The details of this technique and results are presented in \cite{laser}.

\section{Background reduction}
One strength of the SPC is that it allows for simple construction using solely radiopure materials. NEWS-G built the new 140-cm in diameter detector for SNOLAB out of 4N (99.99\% pure) Aurubis copper \cite{filter}. Pure copper has no long-lived radioisotopes making it an ideal construction material for a NEWS-G detector. Recent measurements demonstrated that the 4N copper contained unacceptable amounts of $^{210}$Po and $^{210}$Pb coming from the same decay chain as $^{210}$Rn~\cite{koba}, reducing the sensitivity of the experiment, due to a contribution of 4.6 dru below 1 keV in the background rate from decays of $^{210}$Pb and $^{210}$Bi, an order of magnitude larger contribution than any other background. Thus a 500 {\textmu}m layer of ultra-pure copper was electroplated onto the detector inner surface (with a rate of 0.036 mm/day), which was estimated to reduce this background to 2.0 dru below 1 keV. The electroplating was carried out in the underground laboratory at LSM where the detector vessel was stored. 

\section{Summary}
The NEWS-G SPC filled with light gases provides a window to search the search for light DM in the 0.1-10 GeV$^{2}$ range. Recent results from SEDINE provide competitive constraints on the WIMP-nucleon cross section below 1 GeV/c$^{2}$. In the future operation of the 140-cm low background SPC at SNOLAB, with novel sensor technology and detector monitoring will extend the sensitivity by orders of magnitudes and could shed light on the nature of dark matter.

\section*{Acknowledgements} \label{sec:acknowledgements}
This project has received funding from the European Union's Horizon 2020 research and innovation programme under the Marie
Sk\l{}odowska-Curie grant agreement no 841261.

\end{document}